# Assessment of Clonal Hematopoiesis of Indeterminate Potential and Future Cardiomyopathy from Cardiac Magnetic Resonance Imaging using Deep Learning in a Cardio-oncology Population

Jiarui Xing*[a], Sangeon Ryu*[a], Shawn Ahn[b], Jeacy Espinoza[a], James L. Cross[c], Stephanie Halene[d], James S. Duncan[a], Alokkumar Jha**[b], Jennifer M Kwan**[b], Nicha C. Dvornek**[a]

[a] Department of Radiology & Biomedical Imaging, Yale School of Medicine, New Haven, USA

[b] Department of Surgery, University of Pennsylvania

[c] Section of Cardiovascular Medicine, Yale School of Medicine, New Haven, USA

[d] Section of Hematology, Yale School of Medicine, New Haven, USA

*Co-first authors

*Co-corresponding authors

**Keywords:** Deep learning, clonal hematopoiesis of indeterminate potential, cardiovascular disease, cardiac MRI, cardio-oncology

## Abstract

Background

We propose deep learning methods to non-invasively detect clonal hematopoiesis of indeterminate potential (CHIP) using late gadolinium enhancement cardiac magnetic resonance (LGE CMR) imaging, providing a radiogenomic approach to identify those who have a higher likelihood of having this novel cardiovascular risk factor. Additionally, although CHIP is a risk factor for cardiomyopathy, identifying the subset of CHIP-positive patients have an even higher risk for this adverse cardiovascular outcome using imaging features can guide monitoring and management of these patients.

Methods

We analyzed 136 cardio-oncology patients who underwent LGE CMR imaging, with 152 scanning series included. A deep learning model using ResNet-18 architecture was developed to predict CHIP status from multi-view LGE images using 5-fold cross-validation. The model's ability to predict future cardiomyopathy specifically within the CHIP-positive population was subsequently evaluated to test the framework's utility for risk stratification.

Results

Our deep learning model achieved robust performance for CHIP prediction with an AUC of 0.71 and accuracy of 73%, demonstrating the feasibility of non-invasive CHIP detection from cardiac imaging. In the risk stratification task for CHIP-positive patients, the model successfully predicted future cardiomyopathy with an AUC of 0.87 and accuracy of 81%, demonstrating that LGE signatures can effectively identify the subset of CHIP carriers on a trajectory toward heart failure.

Conclusions

LGE-CMR demonstrates promise for cardiovascular outcome prediction through deep learning approaches. We demonstrated subsequent cardiomyopathy risk prediction in CHIP-positive patients, creating a CMR-based screening system for high-risk cardiovascular populations. Our results support the feasibility of using LGE imaging for CHIP identification and subsequent cardiomyopathy risk assessment, representing an important step toward accessible, imaging-based precision medicine in high-risk cardiovascular populations.

## 1. Introduction

Clonal hematopoiesis of indeterminate potential (CHIP) is an age-related premalignant condition, characterized by the presence of clonally expanded hematopoietic stem cells caused by a leukemogenic mutation in individuals without evidence of hematologic malignancy [1]. CHIP is an independent risk factor for cardiovascular diseases (CVDs), such as atherosclerosis, myocardial infarction, and congestive heart failure [2]. We previously showed that CHIP is more prevalent in cancer patients compared to noncancer patients and CHIP is an independent risk factor for development of cardiomyopathy in this cohort [3]. Cardiovascular disease is one of the leading causes of morbidity and mortality worldwide and cardiac MRI is routinely done in cancer patients to evaluate for cardiotoxicity; thus, being able to identify patients with higher likelihood of CHIP using CMR can help tailor closer monitoring strategies and more aggressive risk factor reduction. Further, although CHIP independently increases the risk of heart disease and heart failure, not all CHIP patients develop these adverse cardiovascular events [4]. Thus, the use of machine learning approaches can potentially identify imaging features that can risk stratify those who may develop CVD, particularly cardiomyopathy amongst CHIP patients.

Our preliminary data shows that CHIP is associated with increased fibrosis pathways in human engineered heart tissue and increased burden of LGE in CMR imaging [5]. LGE is the method of choice for detecting myocardial fibrosis in magnetic resonance imaging. Of note, most cardio-oncology patients undergo routine clinical CMR to evaluate cardiotoxicity. Thus, we propose a deep learning framework with a dual clinical purpose: (1) screening: to non-invasively detect probable CHIP carriers using LGE images, and (2) risk stratification: to analyze these imaging features within CHIP-positive patients to predict future cardiomyopathy. We sought to explore whether a CMR-based framework using LGE signatures could indicate CHIP status and predict future cardiomyopathy development in our cardio-oncology population.

## 2. Methods

### 2.1 Study Subjects and Data

This prospective cohort study enrolled 145 patients with cancer receiving chemotherapy from the cardio-oncology service at Yale-New Haven Hospital between 2019 and 2023 under an institutional review board-approved protocol. All participants provided informed consent prior to enrollment. The cohort had a mean age of 65 ± 13 years, and 66% were female. Comorbidities included arrhythmias (24%), coronary artery disease (61%), diabetes (13%), cardiomyopathy (41%), and hypertension (46%). The most common cancer types were breast (41%) and lung (19%), followed by genitourinary (10%), skin (9%), and renal (6%). Patients received therapies including anthracyclines (31%), HER2 inhibitors (16%), and immune checkpoint inhibitors (49%).

Inclusion criteria comprised adults (≥18 years) scheduled for cancer therapy with baseline CMR and follow-up cardiac imaging (CMR or echocardiography). Exclusion criteria included contraindications to CMR or incomplete imaging data. Peripheral blood samples were collected at baseline for clonal hematopoiesis of indeterminate potential (CHIP) assessment using targeted next-generation sequencing of 74 genes recurrently mutated in hematologic malignancies (median depth 1000x; variant allele frequency threshold 2%). CHIP was detected in 41.5% of patients in the analytic cohort. The most common mutated genes were DNMT3A (40%), TET2 (18%), and PPM1D (15%).

Cardiac function was assessed using 1.5T or 3T CMR scanners with standardized protocols including cine steady-state free precession sequences for volumetric analysis and LGE imaging performed 10 minutes after intravenous gadolinium administration (0.15-0.2 mmol/kg).

The final analysis included 152 CMR scanning series from 136 patients after excluding those with incomplete LGE sequences or inadequate image quality. Cardiotoxicity (cardiomyopathy) was defined as a left ventricular ejection fraction (LVEF) decline of ≥10% to an absolute value <50% per current cardio-oncology guidelines. For the cardiomyopathy prediction task, we identified a global subset of 131 CMR series from 118 patients who had valid longitudinal follow-up data. To evaluate risk stratification specifically among susceptible individuals, our primary analysis focused on the CHIP-positive sub-cohort, which consisted of 52 serial scans from 49 patients. Within this high-risk subset, 16 patients (30.8%) developed cardiomyopathy, while 36 (69.2%) did not.

### 2.2 CNN Model

We developed a unified deep learning framework to address two clinical tasks: (1) CHIP Detection, and (2) Risk Stratification via cardiomyopathy prediction. We trained separate models with identical architectures for each task. The model was based on a ResNet [6] architecture with 18 convolutional layers pretrained on ImageNet [7] and adapted for binary classification. To accommodate the multi-view CMR data, we modified the input layer to accept 4-channel inputs, where each channel corresponds to one of the four CMR views (short axis (SAS), 4 chambers, vertical long axis (VLA), and LVOT), together forming an image set. This approach allows the model to leverage the pretrained features while processing all views simultaneously through a single network. The final fully connected layer was replaced to output the probability of CHIP based on the 4-view CMR image set.

### 2.3 Training and Evaluation

The model was trained using a 5-fold cross validation framework in order to assess the performance of the model. Each fold contained between 25 and 30 patients; we ensured that all CMR images belonging to the same patient were in the same fold. Each of the 5 folds were used once as a test set, while the other 4 folds were combined to be the training set. In addition to standard image data augmentation techniques including random rotation, translation and contrast adjustment, as each view may include multiple image slices, random combinations of the 4 views were used to augment the number of samples. The model for each fold was trained using binary cross-entropy loss for 300 epochs with early stop strategy and then evaluated on the test set. The evaluation profiles were then combined to give an overview of the model architecture's performance in the binary classification task using receiver operating characteristic (ROC) curve analysis, including measuring area under the curve (AUC) and accuracy.

The model performs classification at the "image-level," classifying each 4-view CMR image set into binary categories (CHIP vs. no-CHIP for CHIP prediction; cardiomyopathy vs. no-cardiomyopathy for cardiomyopathy prediction). To extend this to patient-level classification, we combined predictions from all image sets for each patient using a ratio-thresholding approach, where a patient was considered positive if the proportion of positive image-level predictions exceeded 40%. We also applied model interpretation algorithm [8] to highlight CMR regions that contributed most to the prediction, allowing us to visualize potential imaging biomarkers and patterns associated with CHIP.

## 3. Results

### 3.1 Model Performance

Using our patient-level prediction strategy with ratio-thresholding aggregation, our deep learning model achieved robust performance for CHIP prediction with an AUC of 0.71, accuracy of 73%, sensitivity of 0.65 and specificity of 0.65, demonstrating the feasibility of non-invasive CHIP detection from cardiac imaging. In CHIP-positive patients, direct LGE-based cardiomyopathy prediction showed high performance with an AUC of 0.87, accuracy of 81%, sensitivity of 0.61 and specificity of 0.94, suggesting that CMR features can also help stratify cardiovascular risk within the CHIP population. The ROC curves for both tasks are presented in the graphical abstract. In the global population, LGE was also able to predict who would develop CM and achieved AUC of 0.70, accuracy of 85%, sensitivity of 0.46 and specificity of 0.95. Representative heatmap visualizations are shown in Figure 1. We present these heatmaps restricted to the myocardium (via masking) to demonstrate the model's specific focus on cardiac tissue. The highlighted regions overlap with areas of late gadolinium enhancement, indicating that myocardial scar contributed to CHIP-positive predictions.

### 3.2 Independence from Demographic Confounders

To rule out the possibility of the model relying on potential demographic confounders, we explicitly investigated the influence of age and Immune Checkpoint Inhibitor (ICI) usage. We compared the distributions of these variables between cohorts, noting that CHIP-positive patients were slightly older (65.2 ± 16.6 vs. 62.3 ± 13.3 years, $p=0.313$) and had a higher rate of ICI usage (61.7% vs. 43.3%, $p=0.081$). Although these differences did not reach statistical significance in this analytic cohort, we performed a two-stage independence analysis to ensure our model relies on genuine imaging features rather than these potential demographic proxies.

First, we trained "tabular-only" baselines including a logistic regression model and a 3-layer fully connected network (identical to the proposed network's classification head) using only age and ICI status as inputs. As shown in Table 1, these models failed to distinguish CHIP status, achieving accuracies of 57.0% and 59.8% (AUC 0.65–0.68), which is significantly lower than the performance of our image-based model (Accuracy 73%, AUC 0.71). This performance gap indicates that demographic variables alone lack the signal required for robust prediction.

We further investigated this by training a hybrid "late-fusion" model that concatenates the latent features from our proposed ResNet-18 image encoder with normalized age and ICI values before the final classification layer. We applied gradient-based saliency, feature ablation, and permutation importance to quantify the contribution of each input type. As detailed in Table 2, the model assigned >99% of its gradient sensitivity to the image features, while the contribution of Age and ICI was negligible (<0.2%). Consistently, ablating (zeroing) or permuting the demographic inputs resulted in zero drop in model performance (change in AUROC = 0.00), confirming that the LGE-CMR imaging features provide a distinct and independent prognostic signal.

## 4. Conclusions

In conclusion, we developed deep learning models for both CHIP detection (AUC 0.71, accuracy 73%) and cardiomyopathy prediction in CHIP-positive patients (AUC 0.87, accuracy 81%) using multi-view LGE CMR in cardio-oncology patients. We were also able to predict who would develop future cardiomyopathy (with AUC of 0.70, accuracy of 85%) for the entire cohort. Our results establish a comprehensive LGE CMR based diagnostic framework that can non-invasively identify high-risk patients and predict cardiovascular outcomes. This approach leverages routine clinical imaging already performed in cardio-oncology patients, representing an important step toward accessible, imaging-based precision medicine in high-risk cardiovascular populations. Future work will extend validation of our approach for both CHIP detection and cardiomyopathy prediction on large public datasets (e.g., TOPMed [9] and UK Biobank [10]) and further apply deep model interpretation techniques to identify CMR biomarkers for CHIP as well as imaging features that can predict adverse cardiovascular outcomes in CHIP patients with or without cancer.

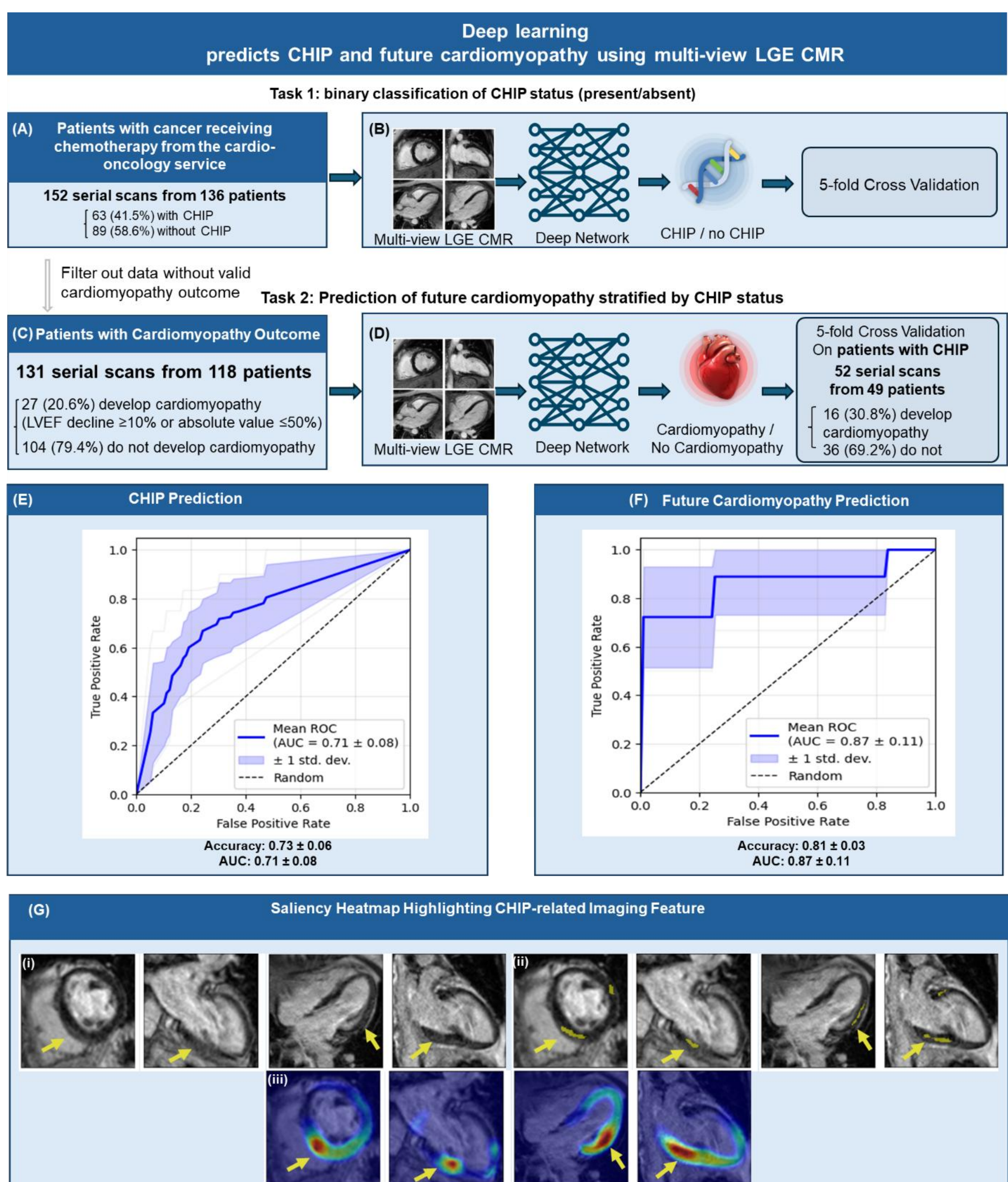

Figure 1. Deep learning framework and performance. (A–B) Workflow for Task 1: Non-invasive CHIP classification from multi-view LGE CMR. (C–D) Workflow for Task 2: Cardiomyopathy risk stratification specifically within the CHIP-positive cohort. (E) ROC curve for Task 1 (CHIP detection, AUC = 0.71). (F) ROC curve for Task 2 (Cardiomyopathy prediction in CHIP+ patients, AUC = 0.87). (G) Representative saliency heatmaps restricted to the myocardium. (i) LGE magnitude images with yellow arrows indicating scar; (ii) Manual scar annotations; (iii) Grad-CAM visualizations showing the model correctly focusing on myocardial scar/fibrosis to generate predictions (red = high contribution. restricted to the myocardium region).

| Model Input | Architecture | Accuracy | AUROC |
|---|---|---|---|
| Age + ICI | Logistic Regression | 0.570 ± 0.089 | 0.652 ± 0.089 |
| Age + ICI | Fully Connected Network | 0.598 ± 0.032 | 0.685 ± 0.069 |
| Image (LGE) | Proposed ResNet-18 | 0.730 ± 0.060 | 0.710 ± 0.080 |

Table 1. Comparison of predictive performance between demographic-only baselines and the proposed image-based model. The age + ICI baselines (LR and FCN) perform significantly worse than the image-based model, demonstrating that Age and ICI status alone are insufficient to explain the model's predictive success.

| **Input Feature** | **Gradient Contribution** | **Ablation Impact (ΔAUROC)** | **Permutation Impact (ΔAUROC)** |
|---|---|---|---|
| **Imaging Features** | **99.87% ± 0.04%** | **-0.085 ± 0.209** | -- |
| **Age** | 0.00% | 0 | 0.000 ± 0.000 |
| **ICI** | 0.12% ± 0.04% | 0 | 0.000 ± 0.003 |

Table 2. Feature importance analysis of the Hybrid (Late-Fusion) model. Gradient contribution indicates the sensitivity of the output to the input. Ablation and Permutation metrics measure the performance impact when specific features are removed (ablation) or randomized (permutation). A value near zero indicates that permutating or removing that feature had no effect on the model's accuracy, proving the model was not relying on it.

## References

[1] C.S. Marnell, A. Bick, P. Natarajan, Clonal hematopoiesis of indeterminate potential (CHIP): Linking somatic mutations, hematopoiesis, chronic inflammation and cardiovascular disease, J Mol Cell Cardiol 161 (2021) 98–105. https://doi.org/10.1016/j.yjmcc.2021.07.004.

[2] L. Mooney, C.S. Goodyear, T. Chandra, K. Kirschner, M. Copland, M.C. Petrie, N.N. Lang, Clonal haematopoiesis of indeterminate potential: intersections between inflammation, vascular disease and heart failure, Clin Sci (Lond) 135 (2021) 991–1007. https://doi.org/10.1042/CS20200306.

[3] E. Leveille, R. Jaber Cheheyeb, C. Matute-Martinez, N.W. Chen, R. Jayakrishnan, A. Christofides, D. Lin, Y. Im, G. Biancon, J. VanOudenhove, S. Halene, J.M. Kwan, Clonal Hematopoiesis Is Associated With Cardiomyopathy During Solid Tumor Therapy, JACC CardioOncol 6 (2024) 605–607. https://doi.org/10.1016/j.jaccao.2024.05.013.

[4] L.D. Weeks, A. Niroula, D. Neuberg, W. Wong, R.C. Lindsley, M. Luskin, N. Berliner, R.M. Stone, D.J. DeAngelo, R. Soiffer, M.M. Uddin, G. Griffin, C. Vlasschaert, C.J. Gibson, S. Jaiswal, A.G. Bick, L. Malcovati, P. Natarajan, B.L. Ebert, Prediction of risk for myeloid malignancy in clonal hematopoiesis, NEJM Evid 2 (2023). https://doi.org/10.1056/evidoa2200310.

[5] H. Kaur, S. Halder, J. Espinoza, N. Chen, J. VanOudenhove, G. Biancon, S. Gu, R. Chakraborty, K. Jain, J. Park, Y. Qyang, E. Ibrahim, S. Halene, J. Hwa, S. Campbell, A. Jha, J. Kwan, Abstract 4144255: Treatment of Human Engineered Heart Tissue with Serum from Cancer Patients with Clonal Hematopoiesis of Indeterminate Potential Cytokine Leads to Impaired Function, Circulation 150 (2024) A4144255–A4144255. https://doi.org/10.1161/circ.150.suppl_1.4144255.

[6] K. He, X. Zhang, S. Ren, J. Sun, Deep Residual Learning for Image Recognition, (2015). https://doi.org/10.48550/arXiv.1512.03385.

[7] J. Deng, W. Dong, R. Socher, L.-J. Li, K. Li, L. Fei-Fei, ImageNet: A large-scale hierarchical image database, in: 2009 IEEE Conference on Computer Vision and Pattern Recognition, 2009: pp. 248–255. https://doi.org/10.1109/CVPR.2009.5206848.

[8] R.R. Selvaraju, M. Cogswell, A. Das, R. Vedantam, D. Parikh, D. Batra, Grad-CAM: Visual Explanations from Deep Networks via Gradient-Based Localization, in: 2017 IEEE International Conference on Computer Vision (ICCV), 2017: pp. 618–626. https://doi.org/10.1109/ICCV.2017.74.

[9] D. Taliun, D.N. Harris, M.D. Kessler, J. Carlson, Z.A. Szpiech, R. Torres, S.A.G. Taliun, A. Corvelo, S.M. Gogarten, H.M. Kang, A.N. Pitsillides, J. LeFaive, S. Lee, X. Tian, B.L. Browning, S. Das, A.-K. Emde, W.E. Clarke, D.P. Loesch, A.C. Shetty, T.W. Blackwell, A.V. Smith, Q. Wong, X. Liu, M.P. Conomos, D.M. Bobo, F. Aguet, C. Albert, A. Alonso, K.G. Ardlie, D.E. Arking, S. Aslibekyan, P.L. Auer, J. Barnard, R.G. Barr, L. Barwick, L.C. Becker, R.L. Beer, E.J. Benjamin, L.F. Bielak, J. Blangero, M. Boehnke, D.W. Bowden, J.A. Brody, E.G. Burchard, B.E. Cade, J.F. Casella, B. Chalazan, D.I. Chasman, Y.-D.I. Chen, M.H. Cho, S.H. Choi, M.K. Chung, C.B. Clish, A. Correa, J.E. Curran, B. Custer, D. Darbar, M. Daya, M. de Andrade, D.L. DeMeo, S.K. Dutcher, P.T. Ellinor, L.S. Emery, C. Eng, D. Fatkin, T. Fingerlin, L. Forer, M. Fornage, N. Franceschini, C. Fuchsberger, S.M. Fullerton, S. Germer, M.T. Gladwin, D.J. Gottlieb, X. Guo, M.E. Hall, J. He, N.L. Heard-Costa, S.R. Heckbert, M.R. Irvin, J.M. Johnsen, A.D. Johnson, R. Kaplan, S.L.R. Kardia, T. Kelly, S. Kelly, E.E. Kenny, D.P. Kiel, R. Klemmer, B.A. Konkle, C. Kooperberg, A. Köttgen, L.A. Lange, J. Lasky-Su, D. Levy, X. Lin, K.-H. Lin, C. Liu, R.J.F. Loos, L. Garman, R. Gerszten, S.A. Lubitz, K.L. Lunetta, A.C.Y. Mak, A. Manichaikul, A.K. Manning, R.A. Mathias, D.D.

McManus, S.T. McGarvey, J.B. Meigs, D.A. Meyers, J.L. Mikulla, M.A. Minear, B.D. Mitchell, S. Mohanty, M.E. Montasser, C. Montgomery, A.C. Morrison, J.M. Murabito, A. Natale, P. Natarajan, S.C. Nelson, K.E. North, J.R. O'Connell, N.D. Palmer, N. Pankratz, G.M. Peloso, P.A. Peyser, J. Pleiness, W.S. Post, B.M. Psaty, D.C. Rao, S. Redline, A.P. Reiner, D. Roden, J.I. Rotter, I. Ruczinski, C. Sarnowski, S. Schoenherr, D.A. Schwartz, J.-S. Seo, S. Seshadri, V.A. Sheehan, W.H. Sheu, M.B. Shoemaker, N.L. Smith, J.A. Smith, N. Sotoodehnia, A.M. Stilp, W. Tang, K.D. Taylor, M. Telen, T.A. Thornton, R.P. Tracy, D.J. Van Den Berg, R.S. Vasan, K.A. Viaud-Martinez, S. Vrieze, D.E. Weeks, B.S. Weir, S.T. Weiss, L.-C. Weng, C.J. Willer, Y. Zhang, X. Zhao, D.K. Arnett, A.E. Ashley-Koch, K.C. Barnes, E. Boerwinkle, S. Gabriel, R. Gibbs, K.M. Rice, S.S. Rich, E.K. Silverman, P. Qasba, W. Gan, G.J. Papanicolaou, D.A. Nickerson, S.R. Browning, M.C. Zody, S. Zöllner, J.G. Wilson, L.A. Cupples, C.C. Laurie, C.E. Jaquish, R.D. Hernandez, T.D. O'Connor, G.R. Abecasis, Sequencing of 53,831 diverse genomes from the NHLBI TOPMed Program, Nature 590 (2021) 290–299. https://doi.org/10.1038/s41586-021-03205-y.

[10] C. Bycroft, C. Freeman, D. Petkova, G. Band, L.T. Elliott, K. Sharp, A. Motyer, D. Vukcevic, O. Delaneau, J. O'Connell, A. Cortes, S. Welsh, A. Young, M. Effingham, G. McVean, S. Leslie, N. Allen, P. Donnelly, J. Marchini, The UK Biobank resource with deep phenotyping and genomic data, Nature 562 (2018) 203–209. https://doi.org/10.1038/s41586-018-0579-z.